\documentstyle[preprint,prd,aps,epsf]{revtex}
\tightenlines

\begin{document}
\draft

\title{Quantum Fluctuations of Effective Fields \\
and the Correspondence Principle}

\author{Kirill~A.~Kazakov\thanks{E-mail address: $kirill@theor.phys.msu.su$}}

\address{Department of Theoretical Physics,
Physics Faculty,\\
Moscow State University,
$117234$, Moscow, Russian Federation}

\maketitle

\begin{abstract}
The question of Bohr correspondence in quantum field theory
is considered from a dynamical point of view. It is shown that
the classical description of particle interactions is inapplicable
even in the limit of large particles' masses because of finite
quantum fluctuations of the fields produced. In particular, it is found
that the relative value of the root mean square fluctuation of the
Coulomb and Newton potentials of a massive particle is equal to
$1/\sqrt{2}\,.$ It is shown also that in the case of a macroscopic
body, the quantum fluctuations are suppressed by a factor $1/\sqrt{N}\,,$
where $N$ is the number of particles in the body. An adequate macroscopic
interpretation of the correspondence principle is given.
\end{abstract}

\section{Introduction}

The Bohr correspondence principle is one of the basic principles
underlying quantum theory. An essential of this principle is the formulation
of quasi-classical transition from the quantum theory to its classical
original, {\it i.e.,} indication of the conditions under which a given system
can be considered classically.

Identification of the quasi-classical conditions is a twofold problem.
Roughly speaking, it consists of the kinematical and dynamical parts.
The former is concerned with the motion of a system, while the latter with
its interactions. From the point of view of quantum kinematics,
determination of the quasi-classical conditions is a quantum mechanical
problem. The motion of a system can be considered classically if
\begin{eqnarray}\label{gg}
\bar{S}/\hbar\gg 1,
\end{eqnarray}
\noindent
where $\bar{S}$ is the characteristic value of the system action.
This is the case, for instance, for a sufficiently heavy particle.
On the other hand, the dynamical part is essentially a quantum field
problem. It requires examination of the fundamental interactions of
the system.

At first sight, the above division of the whole problem into two parts
is artificial. In quantum electrodynamics, for instance, the interaction
of two charged particles takes the classical form of the Coulomb law if the
particles' masses tend to infinity, since all radiative corrections to the
electromagnetic form factors disappear in this limit. This is not the case,
however, in quantum gravity, where the radiative corrections do not disappear
in the large mass limit, because the value of the particle mass determines
the strength of its gravitational interactions. In fact, the relative value
of the logarithmic contribution (which is of the order $\hbar$) to the
gravitational form factors of the scalar particle is independent of the
scalar particle mass \cite{donoghue}. Moreover, there are contributions of
the order $\hbar^0$ which are proportional to the particle mass and have the
form of the ordinary post-Newtonian corrections \cite{iwasaki}.

The question of the correspondence between classical and quantum theories
of gravity was considered in detail in Refs.~\cite{kazakov1,kazakov2}, where
it was shown that the correct correspondence can only be established in the
{\it macroscopic} limit. Namely, the following formulation of the
correspondence principle was suggested: the effective gravitational field
produced by a macroscopic body of mass $M$ consisting of $N$ particles turns
into the corresponding solution of the classical equations in the limit
$M\to\infty,$ $N\to\infty.$ This interpretation is underlined by an
observation that the $n$-loop radiative contribution to the post-Newtonian
correction of a given order to the gravitational field
of a body with mass $M,$ consisting of $N = M/m$
elementary particles with mass $m,$ contains an extra factor of $1/N^n$
in comparison with the corresponding tree contribution. As this fact
is central in what follows, let us take as an example the first
post-Newtonian approximation. In this case, there are two types of
contributions to the gravitational field of the body. The first is the
usual post-Newtonian correction predicted by general relativity, reproduced
in quantum theory by the tree diagrams {\it bilinear} in the energy-momentum
tensor $T^{\mu\nu}$ of the particles, see Fig.~\ref{fig1}(a). The second
is the one-loop contribution shown in Fig.~\ref{fig1}(b). Since this diagram
has two particle operators attached, it is only {\it linear} in $T^{\mu\nu}$.
Therefore, when evaluated between $N$-particle states, the former is
proportional to $(m \cdot N) \cdot (m\cdot N) = M^2,$ while the latter,
to $m^2 \cdot N = M^2/N.$


Thus, in the case of gravity, classical consideration of both the kinematics
and dynamics of a system is justified only for macroscopic systems. This
property formally displays the gravitational interaction as an exception
among other fundamental interactions. The aim of this paper is to show
that despite this seemingly natural conclusion, the above formulation of
the correspondence principle must be extended to all interactions.
For this purpose we will investigate {\it quantum fluctuations} of the
effective electromagnetic and gravitational fields. Obviously, system
interactions can be considered classically only in the case of vanishing
fluctuation of the field produced by the system. As we will see, this
requirement puts all fundamental interactions on an equal footing.

The paper is organized as follows. A general formula for the
correlation function of arbitrary fields is written down in Sec.~\ref{tree},
which is then used in the investigation of quantum fluctuations
of electromagnetic and gravitational fields in Secs.~\ref{qfe}
and \ref{qfg}, respectively. The results obtained are discussed in
Sec.~\ref{conclude}. Some technical results used in the text are
collected in two appendices.

Condensed notations of DeWitt \cite{dewitt1} are in force throughout
this paper. Also, right and left derivatives with respect to the
fields and the sources, respectively, are used. The dimensional
regularization of all divergent quantities is assumed.

\section{Tree contributions and quantum fluctuations}\label{tree}

Before we proceed to actual calculations, let us consider the question
of quantum fluctuations of the effective fields in more detail.

Let a given system of interacting fields $\varphi_a, a = 1,...,A,$ be
described by the action $S[\varphi].$ In the case when the system
possesses gauge symmetries (which is of primary concern for us), in addition
to the matter and gauge fields the set $\{\varphi\}$ contains a number of
auxiliary fields (the Faddeev-Popov ghosts, fields introducing the
gauge {\it etc.}). Denoting by $j^a$ the source for the field $\varphi_a,$
we write the generating functional of Green functions
\begin{eqnarray}\label{gener}
Z[j] = \int d\varphi\exp\left\{\frac{i}{\hbar}\left(
S[\varphi] + j^a\varphi_a\right)\right\}\,,
\end{eqnarray}
\noindent
where $d\varphi$ stands for the product of all $d\varphi_a(x):$
$$d\varphi = \prod\limits_{a=1}^{A}\prod\limits_{x} d\varphi_a(x)\,.$$

As was mentioned in the Introduction, our aim is to investigate the
correspondence between classical and quantum theories from the point of
view of quantum fluctuations of the effective fields. Thus, we assume
that the kinematics of our system is such that the inequality (\ref{gg})
takes place, so the leading contribution to the functional
integral (\ref{gener}) is determined by the stationary ``point''
$\varphi_a = \varphi^{(0)}_a$ satisfying the classical equations of motion
\begin{eqnarray}\label{class}
\frac{\delta S}{\delta\varphi_a} = - j^a\,.
\end{eqnarray}
\noindent
Therefore, the generating functional of Green functions takes the
following quasi-classical form
\begin{eqnarray}\label{generqc}
Z^{(0)}[j] = \exp\left\{\frac{i}{\hbar}\left(
S[\varphi^{(0)}] + j^a\varphi^{(0)}_a\right)\right\}\,.
\end{eqnarray}
\noindent
With the help of this equation one finds the quasi-classical mean field
\begin{eqnarray}\label{mean}
\langle\varphi_a\rangle^{(0)}
= - i\hbar\frac{\delta\ln Z^{(0)}[j]}{\delta j^a}
= \frac{\delta\varphi^{(0)}_b}{\delta j^a} (- 1)^{P_b}
\left(\frac{\delta S[\varphi]}{\delta\varphi_b}
+ j^b\right)_{\varphi = \varphi^{(0)}} + \varphi^{(0)}_a
= \varphi^{(0)}_a\,,
\end{eqnarray}
\noindent
where $P_a$ is the Grassmann parity of the field $\varphi_a.$
As expected, the mean field coincides at the tree level with the
corresponding classical solution. Now, let us calculate the
correlation function of two fields $\varphi_a(x), \varphi_b(y).$
Using Eqs.~(\ref{class}), (\ref{mean}), one has
\begin{eqnarray}\label{corr}&&
\langle[\varphi_a(x) - \varphi^{(0)}_a(x)]
[\varphi_b(y) - \varphi^{(0)}_b(y)]\rangle^{(0)}
\nonumber\\&&
= - \frac{\hbar^2}{Z^{(0)}[j]}
\frac{\delta^2 Z^{(0)}[j]}{\delta j^a(x) \delta j^b(y)}
- \varphi^{(0)}_a(x)\varphi^{(0)}_b(y)
\nonumber\\&&
= - i\hbar\frac{\delta\varphi^{(0)}_b(y)}{\delta j^a(x)}
= i\hbar(-1)^{P_a}\left[\frac{\delta^2 S[\varphi]}
{\delta\varphi_b(y)\delta\varphi_a(x)}\right]
_{\varphi = \varphi^{(0)}}^{-1}\,.
\end{eqnarray}
\noindent
Equation (\ref{generqc}) represents zero order approximation of
$Z(j)$ in the formal expansion with respect to $\bar{S}/\hbar\,.$
In this approximation, therefore, the right hand side of Eq.~(\ref{corr})
is to be set zero, concluding that all field fluctuations disappear in the
limit $\hbar/\bar{S}\to 0,$ as it should be in classical theory.
Informally, this conclusion is only justified as long as
$\hbar S^{-1}_{,ab}\to 0$ follows from $\hbar/\bar{S}\to 0.$
In quantum mechanics, this is indeed the case.
Our purpose below will be to decide to what extent this is true
in quantum field theory.

\section{Quantum fluctuations of electromagnetic field}\label{qfe}

Let us take the set $\{\varphi\}$ to consist of a single charged scalar
field $\phi$ and electromagnetic field $A_{\mu}.$ The action for this system
\begin{eqnarray}\label{actione}
S[\phi,A] = {\displaystyle\int} d^4 x
\left\{(\partial_{\mu}\phi^* + i e A_{\mu}\phi^*)
(\partial^{\mu}\phi - ie A^{\mu}\phi) -
\left(\frac{m c}{\hbar}\right)^2\phi^*\phi\right\}
- \frac{1}{4}{\displaystyle\int} d^4 x F_{\mu\nu} F^{\mu\nu}\,,
\end{eqnarray}
\noindent
where $e,m$ are the charge and mass of scalar field quanta, and
$F_{\mu\nu} = \partial_{\mu} A_{\nu} - \partial_{\nu} A_{\mu}.$
Indices are raised and lowered with the help of Minkowski metric
$\eta_{\mu\nu} = {\rm diag}\{+1,-1,-1,-1\}.$

As was explained in the Introduction, our aim is to investigate the
properties of particle interactions in the limiting case where
their kinematics can be considered classically. Thus, we assume
the scalar particles sufficiently heavy, so that to neglect
uncertainty in their positions and velocities.

Consider a particle in the system, and let $a({\bf q})$ be its
momentum probability distribution. We are interested in the fluctuations
of electromagnetic field produced by this particle in a given spacetime
point. According to Eq.~(\ref{corr}), correlation of $A_{\mu}, A_{\nu}$
at the same spacetime point $x$ has the form
\begin{eqnarray}\label{corre}
B_{\mu\nu}(x)\equiv\langle[A_{\mu}(x) - A^{(0)}_{\mu}(x)]
[A_{\nu}(x) - A^{(0)}_{\nu}(x)]\rangle^{(0)}
= i\hbar\left[\frac{\delta^2 S[\phi,A]}
{\delta A_{\mu}(x)\delta A_{\nu}(x)}\right]
_{\phi = \phi^{(0)}\atop A = A^{(0)}}^{-1}\,.
\end{eqnarray}
\noindent
To the leading order in the coupling constant $e,$ the right hand side
of Eq.~(\ref{corre}) is represented by the one-loop diagrams of
Fig.~\ref{fig2}. In the tree approximation, the loop occurs because
the operators $A_{\mu}$ and $A_{\nu}$ are both taken in the same
spacetime point.

Even without detailed calculation, it is clear that the diagram
of Fig.~\ref{fig2}(b) turns into zero in the limit $\hbar/\bar{S}\to 0.$
Indeed, taking into account the structure of the $A^2\phi^2$ vertex
[see Eq.~(\ref{actione})], one readily sees that this diagram
$\sim\hbar e^2/m\to 0,$ for $m\to \infty.$

Thus, it remains only to calculate the diagram of Fig.~\ref{fig2}(a).
Its analytical expression in the momentum space\footnote{To simplify
calculations, in this section, we choose the units in which $\hbar = c = 1.$}
\begin{eqnarray}\label{1a}
I_{\mu\nu}^{(a)}(p) = \frac{- i e^2\mu^{\epsilon}}
{\sqrt{2\varepsilon_{\bf q} 2\varepsilon_{\bf q - p}}}
\int\frac{d^{4-\epsilon} k}{(2\pi)^4}
(2 q_{\alpha} - p_{\alpha} + k_{\alpha})
G^{\phi}(q + k) (2 q_{\beta} + k_{\beta}) G^{\alpha\mu}(p + k)
G^{\beta\nu}(k)\,,
\end{eqnarray}
\noindent
where $G^{\phi}$ is the scalar particle propagator,
$$G^{\phi}(k) = \frac{1}{m^2 - k^2}\,,$$ $G_{\mu\nu}$ the photon
propagator,
\begin{eqnarray}\label{photon}
G_{\mu\nu}(k) = \frac{\eta_{\mu\nu}}{k^2}\,,
\end{eqnarray}
\noindent
$\varepsilon_{\bf q} = \sqrt{m^2 + {\bf q}^2}\,,$ $\mu$ arbitrary mass scale,
and $\epsilon = 4 - d,$ $d$ being the dimensionality of spacetime.

The photon propagator is taken in the Feynman gauge. However, it is shown
in Appendix A that a change of the gauge conditions gives rise to terms
$\sim 1/m$ which disappear in the large particle mass limit (classical
particle kinematics).

Introducing the Schwinger parametrization
\begin{eqnarray}\label{schwinger}&&
\frac{1}{k^2} = - \int_{0}^{\infty} dy\exp\{y k^2\}\,,
\qquad\frac{1}{(k + p)^2} = - \int_{0}^{\infty} dx\exp\{x (k + p)^2\}\,,
\nonumber\\&&
\frac{1}{k^2 + 2 (kq)} = - \int_{0}^{\infty} dz \exp\{z [k^2 + 2 (kq)]\}\,,
\end{eqnarray}
the loop integrals are evaluated using
\begin{eqnarray}\label{loopint}&&
\int~d^{d}k \exp\{ k^2 (x + y + z) + 2 k^{\mu} (x p_{\mu}
+ z q_{\mu}) + p^2 x \}
\nonumber\\&&
= i \left(\frac{\pi}{x + y + z}\right)^{d/2}
\exp\left\{\frac{p^2 x y - m^2 z^2}{x + y + z}\right\}\,.
\end{eqnarray}
\noindent
Then, changing the integration variables $(x,y,z)$ to $(t,u,v)$ via
\begin{eqnarray}\label{change}&&
x = \frac{t (1 + t + u) v}{m^2 (1 + \alpha t u)}\,,
\qquad y = \frac{u (1 + t + u) v}{m^2 (1 + \alpha t u)}\,,
\qquad z = \frac{ (1 + t + u) v}{m^2 (1 + \alpha t u)}\,,
\qquad \alpha \equiv - \frac{p^2}{m^2}\,,
\end{eqnarray}
\noindent
integrating $v$ out, subtracting the ultraviolet divergence\footnote{A
consistent treatment of divergences of this kind is the problem of
the renormalization theory of composite operators. We are not going into
details of the subtraction procedure, since the divergence (\ref{dive})
[as well as divergence (\ref{divh}) in the gravitational case below]
does not interfere with the terms of the order $\hbar^0$ we are interested in.}
\begin{eqnarray}\label{dive}
I_{\mu\nu}^{(a){\rm div}} = - \frac{e^2\eta_{\mu\nu}}{128\pi^2
\sqrt{\varepsilon_{\bf q}\varepsilon_{\bf q - p}}}
\frac{1}{\epsilon}\left(\frac{\mu}{m}\right)^{\epsilon}\,,
\end{eqnarray}
\noindent
and setting $\epsilon = 0,$ we obtain
\begin{eqnarray}\label{1a2}&&
I_{\mu\nu} \equiv (I_{\mu\nu}^{(a)}
- I_{\mu\nu}^{(a){\rm div}})_{\epsilon = 0}
= \frac{e^2}{32\pi^2\sqrt{\varepsilon_{\bf q}\varepsilon_{\bf q - p}}}
\int\limits_{0}^{+ \infty}\int\limits_{0}^{+ \infty}d u d t
\frac{1}{2 m^2 D H^3}
\nonumber\\&&
\times\left\{p_{\mu} p_{\nu}\left[2 H^2 - 3 H
+ \frac{2 m^2}{p^2}(D - 1) + 1\right]
+ (2 q_{\mu} q_{\nu} - p_{\nu} q_{\mu} - p_{\mu} q_{\nu}) (2 H - 1)^2
\right\}\,,
\nonumber\\&&
H \equiv 1 + u + t\,, \quad D\equiv 1 + \alpha u t\,.
\end{eqnarray}
\noindent
The remaining $u,t$-integrals are evaluated in Appendix B. Using
Eq.~(\ref{roots}), and retaining only terms finite in the limit
$m\to\infty,$ we find
\begin{eqnarray}\label{1a3}&&
I_{\mu\nu}
= \frac{e^2}{16 m \sqrt{\varepsilon_{\bf q}\varepsilon_{\bf q - p}}}
\frac{q_{\mu} q_{\nu}}{\sqrt{- p^2}}\,.
\end{eqnarray}
\noindent
Taking into account the wave packet spreading, and going back to the
coordinate space we finally arrive at the following expression for
the correlation function
\begin{eqnarray}\label{corref}&&
B_{\mu\nu}(x)
= \frac{e^2}{16 m}\int\frac{d^3 {\bf p}}{(2\pi)^3}
\frac{d^3{\bf q}}{(2\pi)^3}
\frac{e^{- i p x}}
{\sqrt{- p^2}}
\frac{a^*({\bf q}) a({\bf q - p})q_{\mu} q_{\nu}}
{\sqrt{\varepsilon_{\bf q}\varepsilon_{\bf q - p}}}\,,
\end{eqnarray}
\noindent
where $p_0 = \varepsilon_{\bf q} - \varepsilon_{\bf q - p}\,.$

Let us apply this result to the static case to find the fluctuations
of the Coulomb potential. Thus, we take $a({\bf q})$ such that
$$\int\frac{d^3{\bf q}}{(2\pi)^3} a^*({\bf q}) a({\bf q}){\bf q} = 0\,.$$
The probability distribution function $a({\bf q})$ is generally of the form
$$a({\bf q}) = b({\bf q}) e^{i{\bf q}{\bf r}_0},$$ where ${\bf r}_0$
is the mean particle position, and $b({\bf q})$ (which can be taken real)
describes the shape of the wave packet. Since the particle is assumed
sufficiently heavy, one has up to terms $\sim 1/m,$
$\varepsilon_{\bf q - p}\approx \varepsilon_{\bf q} \approx m,$
and therefore, $p_0 \approx 0.$ By the same reason, one can neglect the space
components of the particle momentum in comparison with its time component.
Furthermore, as long as we are concerned with fluctuations of the Coulomb
potential, we can substitute $b({\bf q - p})$
by $b({\bf q}):$ this amounts to neglecting multipole moments of
the charge distribution. Taking all this into account, we rewrite
Eq.~(\ref{corref}) as
\begin{eqnarray}\label{corrq}&&
B_{\mu\nu}(x)
= \frac{e^2}{16}\int\frac{d^3 {\bf p}}{(2\pi)^3}
\frac{e^{i {\bf p} ({\bf x} - {\bf r}_0)}}{|{\bf p}|}
\int\frac{d^3{\bf q}}{(2\pi)^3}
b^2({\bf q})\delta_{0\mu}\delta_{0\nu}
= \delta_{0\mu}\delta_{0\nu}\frac{e^2}{32\pi^2 r^2}\,,
\end{eqnarray}
\noindent
where $r = |{\bf x} - {\bf r}_0|.$ Thus, the root mean square
fluctuation of the static potential $\Phi^{\rm e}\equiv A_{0}$ of a massive
particle is
\begin{eqnarray}\label{rmse}&&
\sqrt{\left\langle \Delta\Phi^{\rm e}(r)^2 \right\rangle}
= \frac{e}{\sqrt{32}\pi r}\,.
\end{eqnarray}
\noindent
This is to be compared with the Coulomb potential
\begin{eqnarray}\label{coulomb}&&
\Phi^{\rm e}(r) = \frac{e}{4\pi r}\,.
\end{eqnarray}
\noindent

\section{Quantum fluctuations of gravitational field}\label{qfg}

In this section, we will investigate quantum fluctuations of the
gravitational field. As in the preceding section, we consider
a system of quantized scalar matter interacting with the quantized
gravitational field. This time the scalar particles are assumed real.
The action of this system\footnote{Our notation is $R_{\mu\nu} \equiv
R^{\alpha}_{~\mu\alpha\nu} =
\partial_{\alpha}\Gamma^{\alpha}_{\mu\nu} - \cdot\cdot\cdot,
~R \equiv R_{\mu\nu} g^{\mu\nu}, ~g\equiv \det g_{\mu\nu},
~g_{\mu\nu} = {\rm sgn}(+,-,-,-).$}
\begin{eqnarray}\label{actionh}
S[\phi,h] = \frac{1}{2}{\displaystyle\int} d^4 x \sqrt{- g}
\left\{g^{\mu\nu}\partial_{\mu}\phi\partial_{\nu}\phi
- \left(\frac{m c}{\hbar}\right)^2\phi^2\right\}
- \frac{c^3}{k^2}{\displaystyle\int} d^4 x \sqrt{-g}R\,,
\end{eqnarray}
\noindent
where $k^2 = 16\pi G,$ $G$ is the Newton gravitational constant,
and $h_{\mu\nu} = g_{\mu\nu} - \eta_{\mu\nu}$ are chosen as the dynamical
variables.

According to Eq.~(\ref{corr}), correlation function of the components
of gravitational field taken at the same spacetime point $x$
\begin{eqnarray}\label{corrh}
B_{\mu\nu,\alpha\beta}(x)\equiv
\langle[h_{\mu\nu}(x) - h_{\mu\nu}^{(0)}(x)]
[h_{\alpha\beta}(x) - h^{(0)}_{\alpha\beta}(x)]\rangle^{(0)}
= i\hbar\left[\frac{\delta^2 S[\phi,h]}
{\delta h_{\mu\nu}(x)\delta h_{\alpha\beta}(x)}\right]
_{\phi = \phi^{(0)}\atop h = h^{(0)}}^{-1}\,.
\end{eqnarray}
\noindent
Calculation of this function follows the same steps as in the electromagnetic
case, but is more tedious. The leading contribution in the large particle
mass limit is again contained in the diagram of Fig.~\ref{fig2}(a), where
the indices attached to the point of observation ($x$) are now replaced
by the pair of double indices $\mu\nu,\alpha\beta.$
Analytically,\footnote{In this section, we choose the units in which
$k = c = \hbar = 1.$}
\begin{eqnarray}\label{1ag}&&
I_{\mu\nu,\alpha\beta}^{(a)}(p) = \frac{- i\mu^{\epsilon}}
{\sqrt{2\varepsilon_{\bf q} 2\varepsilon_{\bf q - p}}}
\int\frac{d^{4-\epsilon} k}{(2\pi)^4}
\left\{\frac{1}{2}W^{\gamma\delta\rho\tau}
(q_{\rho} - p_{\rho}) (k_{\tau} + q_{\tau})
- \frac{m^2}{2}\eta^{\gamma\delta}\right\}
\nonumber\\&&
\times G^{\phi}(q + k)\left\{\frac{1}{2}W^{\sigma\lambda\zeta\xi}
q_{\zeta} (k_{\xi} + q_{\xi})
- \frac{m^2}{2}\eta^{\sigma\lambda}\right\}
G_{\mu\nu\gamma\delta}(k + p) G_{\alpha\beta\sigma\lambda}(k)\,,
\end{eqnarray}
\noindent
where
\begin{eqnarray}
G_{\mu\nu\sigma\lambda}(k) =
\frac{W_{\mu\nu\sigma\lambda}}{k^2}
\end{eqnarray}
\noindent
is the graviton propagator, and
\begin{eqnarray}
W^{\alpha\beta\gamma\delta} =
\eta^{\alpha\beta} \eta^{\gamma\delta}
- \eta^{\alpha\gamma} \eta^{\beta\delta}
- \eta^{\alpha\delta} \eta^{\beta\gamma}\,.
\nonumber
\end{eqnarray}
\noindent
The graviton propagator is taken in the most convenient DeWitt gauge.
As in the case of electrodynamics, a change of the gauge conditions
fixing general covariance gives rise to terms irrelevant in the classical
limit.

The tensor multiplication in Eq.~(\ref{1ag}) is conveniently performed
with the help of the tensor package \cite{reduce} for the REDUCE system

\begin{eqnarray}\label{1ag2}
I^{(a)}_{\mu\nu,\alpha\beta} &=& \frac{- i\mu^{\epsilon}}
{\sqrt{2\varepsilon_{\bf q} 2\varepsilon_{\bf q - p}}}
\int\frac{d^{4-\epsilon} k}{(2\pi)^4}
\frac{1}{k^2}\frac{1}{(k + p)^2}\frac{1}{m^2 - (k + q)^2}
\nonumber\\
&\times&
\{
m^4 \eta_{\mu\nu}\eta_{\alpha\beta}
- 2 m^2 (\eta_{\mu\nu}q_{\alpha} q_{\beta}
+ \eta_{\alpha\beta}q_{\mu} q_{\nu})
- 4 m^2\eta_{\mu\nu}q_{(\alpha} k_{\beta)}
\nonumber\\
&+& 2 m^2 \eta_{\alpha\beta}(p_{(\mu} q_{\nu)} + p_{(\mu} k_{\nu)}
- q_{(\mu} k_{\nu)})
- 4 p_{(\mu} q_{\nu)} q_{(\alpha} k_{\beta)}
\nonumber\\
&-& 4 q_{\alpha} q_{\beta} (p_{(\mu} q_{\nu)}
+ p_{(\mu} k_{\nu)})
- 4 p_{(\mu} k_{\nu)} q_{(\alpha} k_{\beta)}
\nonumber\\
&+& 8 q_{(\mu} q_{\nu} q_{\alpha} k_{\beta)}
+ 4 q_{(\mu} k_{\nu)} q_{(\alpha} k_{\beta)}
+ 4 q_{\mu} q_{\nu} q_{\alpha} q_{\beta}
\}\,,
\end{eqnarray}
\noindent
where $(\mu_1\mu_2\cdot\cdot\cdot\mu_n)$ denotes symmetrization over
indices enclosed in parentheses,
$$(\mu_1\mu_2\cdot\cdot\cdot\mu_n) = \frac{1}{n!}
\sum\limits_{\{i_1 i_2\cdot\cdot\cdot i_n\} =
\atop {\rm perm}\{ 12\cdot\cdot\cdot n\}}
\mu_{i_1}\mu_{i_2}\cdot\cdot\cdot\mu_{i_n}\,.$$
The loop integral in Eq.~(\ref{1ag2}) is again ultraviolet divergent:
\begin{eqnarray}\label{divh}
I^{(a){\rm div}}_{\mu\nu,\alpha\beta} =
\frac{1}
{256\pi^2\sqrt{\varepsilon_{\bf q}\varepsilon_{\bf q - p}}}
\frac{1}{\epsilon}\left(\frac{\mu}{m}\right)^{\epsilon}
(\eta_{\mu\alpha} p_{\nu} q_{\beta}
+ \eta_{\mu\beta} p_{\nu} q_{\alpha}
+ \eta_{\nu\alpha} p_{\mu} q_{\beta}
&+& \eta_{\nu\beta} p_{\mu} q_{\alpha}
\nonumber\\
- \eta_{\mu\alpha} q_{\nu} q_{\beta}
- \eta_{\mu\beta} q_{\nu} q_{\alpha}
- \eta_{\nu\alpha} q_{\mu} q_{\beta}
&-& \eta_{\nu\beta} q_{\mu} q_{\alpha})\,.
\end{eqnarray}
\noindent
Introducing the Schwinger parametrization Eq.~(\ref{schwinger}),
calculating the loop integrals with the help of Eq.~(\ref{loopint}),
subtracting the ultraviolet divergence (\ref{divh}), and changing the
integration variables according to Eq.~(\ref{change}), we
obtain the following expression\footnote{By itself, the right hand side of
Eq.~(\ref{1ag}) is not symmetric with respect to the interchange
$(\mu\nu)\leftrightarrow (\alpha\beta),$ in particular, its divergent part
(\ref{divh}) is not. However, the part remaining finite in the large particle
mass limit, which is contained entirely in the diagram of Fig.~\ref{fig2}(a),
is symmetric with respect to this interchange. For convenience, the
integrand in Eq.~(\ref{1a2g}) is rewritten in an explicitly symmetric form.
See also the footnote 2.}
for $I_{\mu\nu,\alpha\beta}\equiv
(I^{(a)}_{\mu\nu,\alpha\beta}
- I^{(a){\rm div}}_{\mu\nu,\alpha\beta})_{\epsilon = 0},$ after a trivial
integration over $v$:
\begin{eqnarray}\label{1a2g}&&
I_{\mu\nu,\alpha\beta}
= \frac{1}{32\pi^2\sqrt{\varepsilon_{\bf q}\varepsilon_{\bf q - p}}}
\int\limits_{0}^{+ \infty}\int\limits_{0}^{+ \infty}d u d t
\frac{1}{2 m^2 D H^3}
\nonumber\\&&
\times\left\{
2 \eta_{\mu\nu} \eta_{\alpha\beta} m^4 H^2
+ 4 (\eta_{\mu\nu} p_{(\alpha} q_{\beta)}
+ \eta_{\alpha\beta} p_{(\mu} q_{\nu)}) m^2 H (H - 1)
\right.
\nonumber\\&&
\left.
- 4 (\eta_{\mu\nu} q_{\alpha} q_{\beta}
+ \eta_{\alpha\beta} q_{\mu} q_{\nu})m^2 H (H - 1)
- (\eta_{\alpha\beta} p_{\mu} p_{\nu} +
\eta_{\mu\nu} p_{\alpha} p_{\beta}) m^2 H (H - 1)
\right.
\nonumber\\&&
\left.
- 2 (p_{\mu} p_{\nu} p_{(\alpha} q_{\beta)}
+ p_{\alpha} p_{\beta} p_{(\mu} q_{\nu)})
\left[(H - 1)^2 + \frac{2 m^2}{p^2} (D - 1)\right]
\right.
\nonumber\\&&
\left.
+ 2 (p_{\mu} p_{\nu} q_{\alpha} q_{\beta}
+ p_{\alpha} p_{\beta} q_{\mu} q_{\nu}) (H - 1)^2
+ 8 p_{(\nu} q_{\mu)} p_{(\alpha} q_{\beta)}
\left[(H - 1)^2 + \frac{m^2}{p^2} (D - 1) \right]
\right.
\nonumber\\&&
\left.
- 8 (p_{(\mu} q_{\nu)} q_{\alpha} q_{\beta}
+ p_{(\alpha} q_{\beta)} q_{\mu} q_{\nu}) (H - 1)^2
+ 8 q_{\mu} q_{\nu} q_{\alpha} q_{\beta} (H - 1)^2
\right\}\,.
\end{eqnarray}
\noindent
Using Eq.~(\ref{roots}) of Appendix B, we find the leading at
$m\to\infty$ contribution
\begin{eqnarray}&&
I_{\mu\nu,\alpha\beta}
= \frac{(2 q_{\mu} q_{\nu} - m^2\eta_{\mu\nu})
(2 q_{\alpha} q_{\beta} - m^2\eta_{\alpha\beta})}
{64 m \sqrt{\varepsilon_{\bf q}\varepsilon_{\bf q - p}}\sqrt{- p^2}}\,.
\end{eqnarray}
\noindent
If $a({\bf q})$ is the momentum probability distribution for a given
particle, then the correlation function of the fluctuating gravitational
field produced by this particle takes the form, in the ordinary units,
\begin{eqnarray}\label{corrhf}&&
B_{\mu\nu,\alpha\beta}(x)
\nonumber\\&&
= \frac{4 G^2 \pi^2}{m c^2}\int\frac{d^3 {\bf p}}{(2\pi)^3}
\frac{d^3{\bf q}}{(2\pi)^3}
\frac{e^{- i p x}}
{\sqrt{- p^2}}
\frac{a^*({\bf q}) a({\bf q - p})}
{\sqrt{\varepsilon_{\bf q}\varepsilon_{\bf q - p}}}
\left(\frac{2 q_{\mu} q_{\nu}}{c^2} - m^2\eta_{\mu\nu} \right)
\left(\frac{2 q_{\alpha} q_{\beta}}{c^2} - m^2\eta_{\alpha\beta}\right)\,,
\end{eqnarray}
\noindent
where $p_0 = \varepsilon_{\bf q} - \varepsilon_{\bf q - p}\,.$

Following discussion in the preceding section, we apply this result to
the static case. Setting
$\varepsilon_{\bf q - p}\approx \varepsilon_{\bf q} \approx m c^2,$
$p_0 \approx 0,$ we see that only the components with $\mu = \nu,$
$\alpha = \beta$ survive in the limit $m\to\infty,$ being all equal
to each other:
\begin{eqnarray}\label{corrhs}&&
B_{\mu\nu,\alpha\beta}(x)
= \frac{2 G^2 m^2}{r^2 c^4}\delta_{\mu\nu}\delta_{\alpha\beta}\,,
\end{eqnarray}
\noindent
where $r = |{\bf x} - {\bf r}_0|$ is the distance between the observation
point $x$ and the mean particle position ${\bf r}_0.$
In particular, the root mean square fluctuation of $\mu\nu$-component
of the metric
\begin{eqnarray}\label{rmsh}&&
\sqrt{\left\langle \Delta h_{\mu\nu}(r)^2 \right\rangle}
= \delta_{\mu\nu}\frac{\sqrt{2}G m}{r c^2}\,.
\end{eqnarray}
\noindent
The nonrelativistic gravitational potential $\Phi^{\rm g}$ is related
to the $00$-component of metric as $\Phi^{\rm g} = h_{00} c^2/2\,,$
thus,
\begin{eqnarray}\label{rmsp}&&
\sqrt{\left\langle \Delta \Phi^{\rm g}(r)^2 \right\rangle}
= \frac{G m}{\sqrt{2}r}\,.
\end{eqnarray}
\noindent
We see that in both electromagnetic and gravitational cases,
the relative value of the root mean square fluctuation of the
potential is equal to $1/\sqrt{2}\,.$

\section{Discussion and conclusions}\label{conclude}

Equations (\ref{rmse}) and (\ref{rmsp}) describing quantum fluctuations
of the electromagnetic and gravitational fields of a massive particle,
respectively, lead us to an important conclusion that dynamics of an
elementary system which kinematics can be considered classically are
nevertheless essentially quantum. The relative value of the root mean square
fluctuation of the particle potential turns out to be equal to $1/\sqrt{2}$
in both electromagnetic and gravitational cases. Despite the fact that the
uncertainty in the position and velocity of a sufficiently massive particle
can be completely neglected, its interactions remain essentially quantum.

Quantum character of particle interactions makes the classical consideration
inadequate in the case of systems governed by the interaction of the
constituent particles. It must be mentioned in this connection that there is
a deep-rooted belief in the literature that the quantum field description of
interacting remote systems, each of which consists of many particles, is
equivalent to that in which these systems are replaced by elementary particles
with masses and charges equal to the total masses and charges of the systems.
In other words, the familiar notion of a point particle is carried over from
the classical mechanics to the quantum field theory. This point of view is
adhered, for instance, in the classic paper by Iwasaki \cite{iwasaki} where
it is applied to the solar system to calculate the shift of the Mercury
perihelion, considered as a ``Lamb shift''.

\begin{figure}
\epsfxsize=17cm\epsfbox{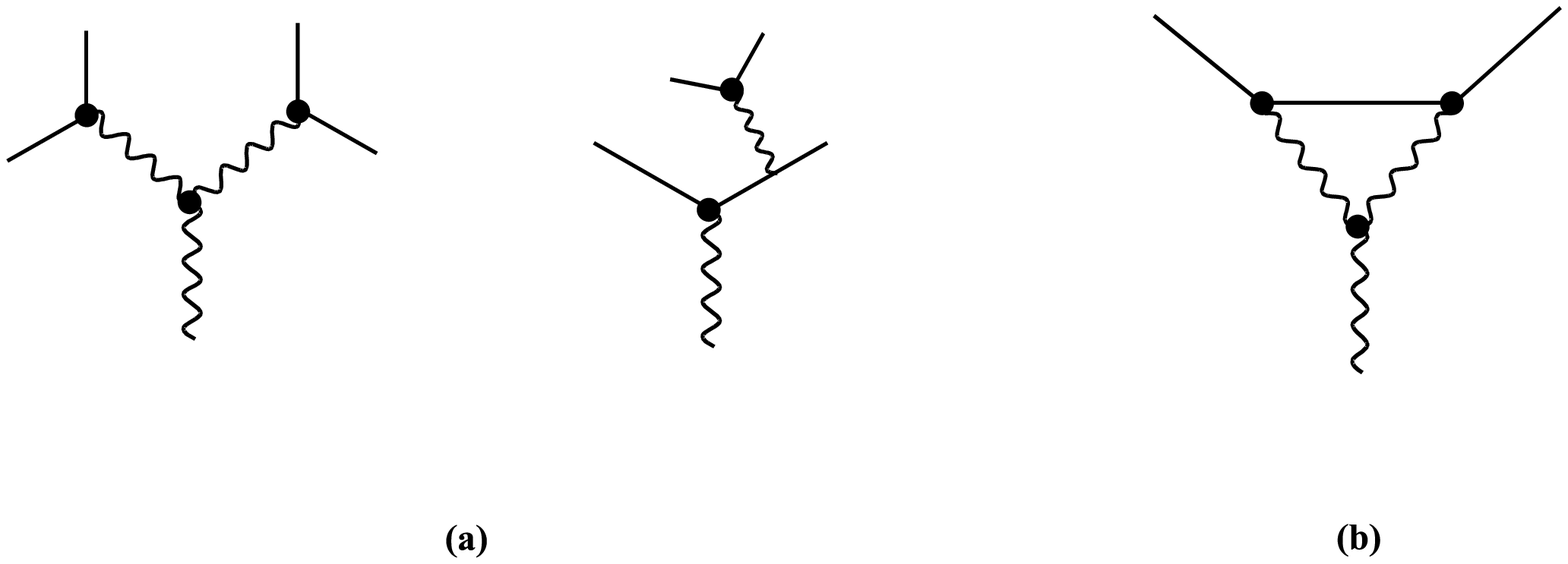}
\vspace{1cm}
\caption{Diagrams contributing to the first post-Newtonian correction.
(a) The tree diagrams occurring because of the gravitational 
self-interaction, and gravitational particle interactions. 
(b) The one-loop radiative correction.  Wavy lines represent
gravitons, solid lines constituent particles.}
\label{fig1}
\end{figure}

\begin{figure}
\epsfxsize=15cm\epsfbox{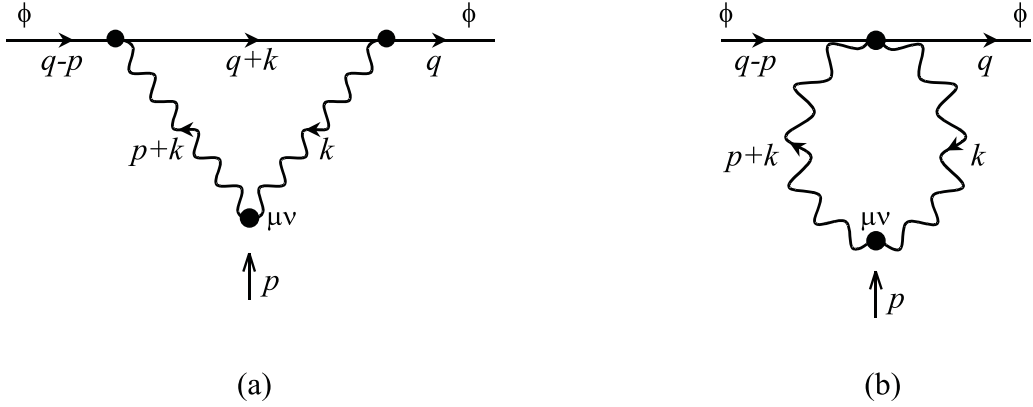}
\vspace{1cm}
\caption{The leading ($\sim e^2$) contribution to the correlation function
$B_{\mu\nu}(x).$}
\label{fig2}
\end{figure}

The Sun and Mercury are regarded in Ref.~\cite{iwasaki} as scalar particles.
As we saw in Sec.~\ref{qfg}, the root mean square fluctuation of the
gravitational potential of the Sun in this case would be $71\%$ of its value.
Fortunately, such fluctuations are not observed in reality. This is because
the Sun is composed of a huge number of elementary particles each of which
contributes to the total gravitational field. To find the resulting quantum
fluctuation of the total field, we turn back to the arguments presented in
the Introduction. As far as the $\phi$-lines are concerned, the diagram
of Fig.~\ref{fig2}(a) is of the same structure as that of Fig.~\ref{fig1}(b).
Therefore, in the case when the gravitational field is produced by a
$N$-particle body, this diagram is proportional to $m^2 N,$ where
$m$ is the mass of a constituent particle. Correspondingly, the root mean
square fluctuation of the potential is proportional to $m\sqrt{N} =
M/\sqrt{N}$ ($M$ is the total mass of the body), while its relative
value, to $1/\sqrt{N}\,.$ If the solar gravitational field is considered,
the quantum fluctuation turns out to be suppressed by a factor of the
order $\sqrt{m_{proton}/M_{\odot}} \approx 10^{-28}\,.$

Another example of attempts to recover the nonlinearity of a classical
theory through the radiative corrections can be found in
Ref.~\cite{donoghue2}. The authors of \cite{donoghue2} claim that the
electromagnetic corrections of the order $e^2$ to the classical
Reissner-Nordstr\"{o}m solution are reproduced by the diagram
of Fig.~\ref{fig1}(b) in which the internal wavy lines correspond to
the virtual photons. However, as we have shown, it is meaningless to try
to establish the correspondence between classical and quantum theories
in terms of {\it elementary} particles, because the quantum fluctuations of
the electromagnetic and gravitational fields produced by such particles are
of the order of the fields themselves. On the other hand, dependence of the
diagram \ref{fig1}(b) on the number of particles is inappropriate to
reproduce the classical physics in the macroscopic limit. This can be shown
using the same argument as in the case of purely gravitational interaction.
Namely, given a body with the total electric charge $Q,$ consisting of
$N = Q/q$ particles with charge $q,$ the contribution of the diagram
\ref{fig1}(b) is proportional to $N\cdot q^2 = Q^2/N\,$ turning into zero
in the macroscopic limit. The relevant contribution correctly reproducing
the $e^2$-correction to the Reissner-Nordstr\"{o}m solution is given by the
tree diagrams of Fig.~\ref{fig1}(a) in which internal wavy lines correspond
to virtual photons.

Thus, we arrive at the conclusion that the requirement of vanishing of the
field fluctuations in the classical limit forces us to extend the macroscopic
formulation of the correspondence principle, suggested in
Ref.~\cite{kazakov1} in the case of gravity, to all interactions.

Finally, in the light of the above discussion, a natural question arises
whether sufficiently massive objects which {\it can} be considered as
elementary particles actually exist in our Universe. An example of such
objects is probably supplied by the black holes. As is well known, black
holes of certain types do behave like normal elementary particles
\cite{wilchek}. Further discussion of this and related issues can be found
in Refs.~\cite{kazakov3,kazakov4}.

\acknowledgments{
I thank Drs. G.~A.~Sardanashvily and P.~I.~Pronin (Moscow State
University) for discussions.}

\begin{appendix}

\section{Gauge independence of the correlation function}

Up to an additive constant, the quantities $\Phi^{\rm e}(r),$
$\Phi^{\rm g}(r)$ determine the potential energy of interacting particles.
Their fluctuations are thus of direct physical significance, and therefore,
expected to be independent of the gauge conditions chosen to fix the gauge
invariance. Let us show that the correlation functions (\ref{corre}),
(\ref{corrh}) remain unchanged under variations of the gauge conditions
indeed. We consider only variations which do not alter the potentials
$\Phi^{\rm e}(r),$ $\Phi^{\rm g}(r)$ themselves, since the question of gauge
independence of their fluctuations would be meaningless otherwise. This
restriction, of course, is not a loss of generality. To allow for completely
arbitrary variations of the gauge conditions, we should have been
dealing with the fluctuations of gauge invariant quantities built from
potentials, rather than the potentials themselves, from the very beginning.

Variations of the gauge conditions which leave the Coulomb potential
unchanged are those satisfying
\begin{eqnarray}\label{epvar}
\delta G_{\mu\nu}(x) = \partial_{\mu} \partial_{\nu} d(x)\,.
\end{eqnarray}
\noindent
Indeed, in this case only
$$\delta A_{\mu}(x) = \int d^4 y~\delta G_{\mu\nu}(x - y)j^{\nu}(y)
= \int d^4 y~\partial^x_{\mu}d(x - y) \partial^y_{\nu} j^{\nu}(y) = 0\,,$$
in view of the current conservation. In the case of gravity,
the corresponding variation of the propagator is more complicated:
\begin{eqnarray}\label{hpvar}
\delta G_{\mu\nu\alpha\beta}(x)
= \partial_{\nu}\partial_{\beta}d_{\mu\alpha}(x)
+ \partial_{\mu}\partial_{\beta}d_{\nu\alpha}(x)
+ \partial_{\nu}\partial_{\alpha}d_{\mu\beta}(x)
+ \partial_{\mu}\partial_{\alpha}d_{\nu\beta}(x)\,,
\qquad d_{\mu\nu} = d_{\nu\mu}\,.
\end{eqnarray}
\noindent
Such a variation induces no variation of the metric (in particular, of
the Newton potential) because of the energy-momentum conservation.

It is not difficult to verify that Eqs.~(\ref{epvar}) and (\ref{hpvar})
can be rewritten as
\begin{eqnarray}\label{epvar1}
\delta G_{\mu\nu} = \partial_{\mu}\xi_{\nu}
\equiv D^{(0)}_{\mu}\xi_{\nu}\,,
\qquad \xi_{\nu} = \partial_{\nu} d\,,
\end{eqnarray}
\noindent
and
\begin{eqnarray}\label{hpvar1}
\delta G_{\mu\nu\alpha\beta}
= \eta_{\mu\gamma}\partial_{\nu}\xi^{\gamma}_{\beta\alpha}
+ \eta_{\nu\gamma}\partial_{\mu}\xi^{\gamma}_{\beta\alpha}
\equiv D^{(0)}_{\mu\nu|\gamma}\xi^{\gamma}_{\beta\alpha}\,,
\qquad \xi^{\gamma}_{\beta\alpha} = 2\partial_{(\beta}d^{\gamma}_{\alpha)}\,,
\end{eqnarray}
\noindent
respectively.
The operators $D^{(0)}$ are nothing but the generators of the gauge
transformations of free electromagnetic and gravitational fields.
To tackle both cases simultaneously, we denote the gauge field
collectively by $Z_{A}$ with a single Latin capital index,
and rewrite Eqs.~(\ref{epvar1}), (\ref{hpvar1}) uniquely as
\begin{eqnarray}\label{unique}
\delta G_{AB}
= D^{(0)}_{A|\gamma}\xi^{\gamma}_{B}
= D^{(0)}_{B|\gamma}\xi^{\gamma}_{A}\,.
\end{eqnarray}
\noindent
(in the electromagnetic case, $\gamma$ takes only one value.)
Using this representation, it is easy to show that the results
(\ref{corref}), (\ref{corrhf}) are invariant with respect to
the gauge transformations (\ref{unique}). We will prove this fact even in a
more general setting when the gauge fields are produced by an arbitrary
species of particles, bosons or fermions, denoted collectively
by $\phi_i,$ $i = 1,...,I.$

It follows from Eq.~(\ref{unique}) that the gauge dependent part of the
gauge field propagators in Fig.~\ref{fig2}(a) is attached to the
matter line through the generator $D^{(0)}\,.$ On the other hand, the
action $S(\phi,Z)$ is gauge invariant
\begin{eqnarray}&&
\frac{\delta S(\phi,Z)}{\delta\phi_i} D_{i|\gamma}
+ \frac{\delta S(\phi,Z)}{\delta Z_{A}}D_{A|\gamma} = 0\,,
\end{eqnarray}
\noindent
where $D_{A|\gamma} = D_{A|\gamma}(Z)$ and
$D_{i|\gamma} = D_{i|\gamma}(\phi)$ are generators of the gauge
transformations of the gauge and matter fields, respectively
[$D_{A|\gamma}(0) \equiv D^{(0)}_{A|\gamma}$].

Differentiating this identity with respect to $\phi_k,$
setting $Z_{A} = 0,$ and taking into account that the
external $\phi$-lines are on the mass shell
$$\frac{\delta S^{(2)}(\phi,0)}{\delta\phi_i} = 0\,,$$ where $S^{(2)}$
denotes the part of $S(\phi,Z)$ bilinear in $\phi,$ the $\phi^2 Z$ vertex
can be rewritten as
\begin{eqnarray}&&
\left.\frac{\delta^2 S^{(2)}(\phi,Z)}{\delta Z_{A}\delta\phi_k}
\right|_{Z=0}D^{(0)}_{A|\gamma} = -
\frac{\delta^2 S^{(2)}(\phi,0)}{\delta\phi_i\delta\phi_k} D_{i|\gamma}\,.
\end{eqnarray}
\noindent Thus, under contraction with the vertex factor, the
$\phi$-particle propagator, $G^{\phi}_{i k},$ satisfying
$$\frac{\delta^2 S^{(2)}(\phi,0)}{\delta\phi_i\delta\phi_k}
G^{\phi}_{k l} = - \delta_{l}^{i}\,,$$
cancels out
\begin{eqnarray}\label{cancel}
G^{\phi}_{k l}\left.\frac{\delta^2 S^{(2)}_{\phi}}{\delta Z_{A}
\delta\phi_k}\right|_{Z=0}D^{(0)}_{A|\gamma} =  D_{l|\gamma}\,.
\end{eqnarray}
\noindent
The root singularity responsible for the $r^{-2}$-behavior of the
correlation function occurs because of the virtual $\phi$-particle
propagation near its mass shell. Thus, the cancellation of the
$\phi$-propagator in the gauge dependent part of the diagram \ref{fig2}(a)
implies the gauge independence of the right hand sides of
Eqs.~(\ref{corref}), (\ref{corrhf}).

\section{Root singularities of Feynman integrals}

The root singularity with respect to the momentum transfer, responsible
for the $r^{-2}$ fall off of the correlation function, is contained
in the integrals
\begin{eqnarray}&&
J_{nm} \equiv \int\limits_{0}^{\infty}\int\limits_{0}^{\infty}
\frac{du dt}{(1 + t + u)^n (1 + \alpha t u)^m}\,,
\nonumber
\end{eqnarray}
encountered in Secs.~\ref{qfe}, \ref{qfg}. It can be extracted as follows.
Consider the auxiliary quantity
\begin{eqnarray}&&
J(A,B) = \int\limits_{0}^{\infty}\int\limits_{0}^{\infty}
\frac{du dt}{(A + t + u) (B + \alpha t u)}\,,
\nonumber
\end{eqnarray}
\noindent
where $A,B>0$ are parameters eventually set equal to 1.
Performing an elementary integration over $u,$ we get
\begin{eqnarray}&&
J(A,B) = \int\limits_{0}^{\infty} dt
~\frac{\ln B - \ln \{\alpha t (A + t)\}}{B - \alpha t (A + t)}\,.
\nonumber
\end{eqnarray}

\begin{figure}
\hspace*{3cm}
\epsfxsize=10cm\epsfbox{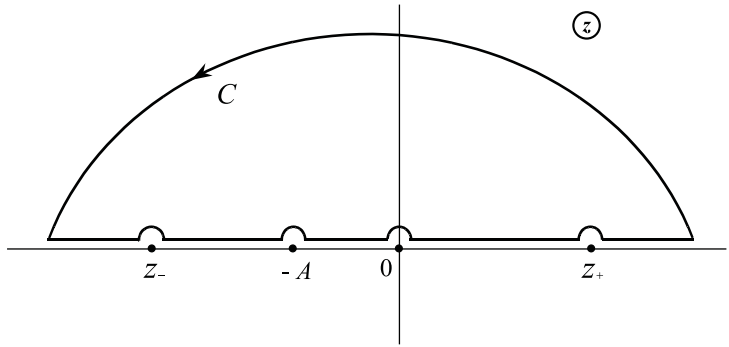}
\vspace{1cm}
\caption{Contour of integration in Eq.~(B1).}
\label{fig3}
\end{figure}

Now consider the integral
\begin{eqnarray}&&
\tilde{J}(A,B) = \oint\limits_{C} dz f(z,A,B),
~~f(z,A,B) = \frac{\ln B - \ln \{\alpha z (A + z)\}}{B - \alpha z (A + z)}\,,
\end{eqnarray}
taken over the contour $C$ shown in Fig.~\ref{fig3}. $\tilde{J}(A,B)$
is zero identically. On the other hand,
\begin{eqnarray}&&
\tilde{J}(A,B)
\nonumber\\&&
= \int\limits^{-A}_{- \infty} dw
~\frac{\ln B - \ln \{\alpha w (A + w)\}}{B - \alpha w (A + w)}
+ \int\limits_{-A}^{0} dw
~\frac{\ln B - \ln \{ - \alpha w (A + w)\} + i\pi}{B - \alpha w (A + w)}
\nonumber\\&&
+ {\rm pv}\int\limits_{0}^{+ \infty} dw
~\frac{\ln B - \ln \{\alpha w (A + w)\} + 2 i\pi}{B - \alpha w (A + w)}
- i\pi \sum\limits_{z_{+},z_{-}} {\rm Res} f(z,A,B)\,,
\nonumber
\end{eqnarray}
\noindent
``pv'' denoting the principal value, and $z_{\pm}$ the poles of the function 
$f(z,A,B),$ 
$$z_{\pm} = - \frac{A}{2} \pm \sqrt{\frac{B}{\alpha} + \frac{A^2}{4}}\ .$$
Change $w \to - A - w$ in the first integral.
A simple calculation then gives
\begin{eqnarray}&&\label{int4}
J(A,B)
= \frac{\pi^2}{2\sqrt{\alpha}} B^{-1/2}
\left(1 + \frac{\alpha A^2}{4 B}\right)^{-1/2}
- \frac{1}{2}\int\limits_{0}^{A} dt
~\frac{\ln B - \ln \{\alpha t (A - t)\}}{B + \alpha t (A - t)}\,.
\end{eqnarray}

The roots are contained entirely in the first term on the right of
Eq.~(\ref{int4}). The integrals $J_{nm}$ are found by repeated
differentiation of Eq.~(\ref{int4}) with respect to $A,B$. Expanding
$(1 + \alpha A^2/4 B)^{-1/2}$ in powers of $\alpha,$
we find the leading terms needed in Eqs.~(\ref{1a2}), (\ref{1a2g})
\begin{eqnarray}&&\label{roots}
J^{{\rm root}}_{11} = \frac{\pi^2}{2\sqrt{\alpha}}\,,
\quad J^{{\rm root}}_{21} = \frac{\pi^2}{8}\sqrt{\alpha}\,,
\quad J^{{\rm root}}_{31} = - \frac{\pi^2}{16}\sqrt{\alpha}\,.
\end{eqnarray}

\end{appendix}

\end{document}